\documentclass[shortnote,twocolumn]{jpsj3}
\usepackage{txfonts}
\usepackage{color}
\usepackage{bm}
\usepackage{here}

\title
{
Strong Suppression of Magnetic Ordering\\in an $S\,{=}\,1/2$ Square-Lattice Heisenberg Antiferromagnet Sr$_2$CuTeO$_6$
}

\author
{
Tomoyuki \textsc{Koga}, Nobuyuki \textsc{Kurita}, and Hidekazu \textsc{Tanaka}\thanks{E-mail address: tanaka@lee.phys.titech.ac.jp}
}

\inst
{
Department of Physics, Tokyo Institute of Technology, Meguro-ku, Tokyo
152-8551
}

\begin{document}
\maketitle

The most appealing physical phenomenon in frustrated quantum magnets is the emergence of nonclassical ground states stabilized by the synergy effect of the spin frustration and the quantum fluctuation. The spin-1/2 square-lattice Heisenberg antiferromagnet (SLHAF) with the nearest-neighbor $J_1$ and next-nearest-neighbor $J_2$ exchange interactions is a typical frustrated quantum magnet, for which energetic theoretical studies have been performed. Most of the theoretical results suggest that the $S\,{=}\,1/2$ $J_1\,{-}\,J_2$ SLHAF exhibits the quantum disordered ground state for ${\alpha}_{\rm c1}\,{<}\,J_2/J_1\,{<}\,{\alpha}_{\rm c2}$ with ${\alpha}_{\rm c1}\,{\simeq}\,0.4$ and ${\alpha}_{\rm c2}\,{\simeq}\,0.6$\cite{Chandra,Dagotto,Figueirido,Read,Igarashi,Einarsson,Zhitomirsky,Bishop,Sirker,Mambrini,Darradi}. However, the nature of the ground state has not been theoretically clarified. The ground states for ${\alpha}_{\rm c1}\,{>}\,J_2/J_1$ and ${\alpha}_{\rm c2}\,{<}\,J_2/J_1$ are the N\'{e}el antiferromagnetic state and the columnar antiferromagnetic state, respectively.

On the experimental side, many materials have been investigated from the viewpoint of the $S\,{=}\,1/2$ $J_1\,{-}\,J_2$ SLHAF\,\cite{Iwanaga,Todate1,Todate2,Vasala,Melzi1,Melzi2,Rosner,Nath,Tsirlin}. However, the quantum disordered ground state has not been observed experimentally. The search for approximate realizations of the $S\,{=}\,1/2$ $J_1\,{-}\,J_2$ SLHAF with the critical parameter range has been continued. 

\begin{figure}[t]
\begin{center}
\hspace{5mm}
\scalebox{0.32}{
\includegraphics{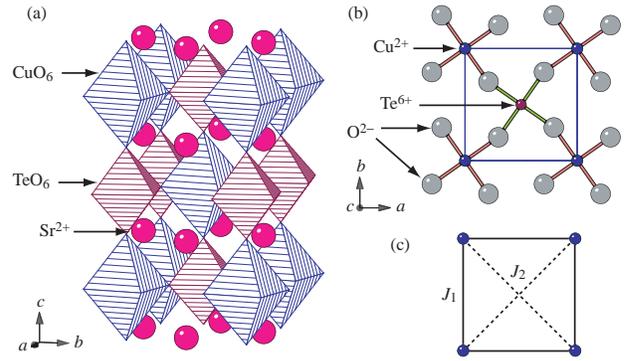}
}
\end{center}
\vspace{-2mm}
\caption{(Color online) (a) Schematic crystal structure of Sr$_2$CuTeO$_6$. The blue octahedron is a CuO$_6$ octahedron with a Cu$^{2+}$ ion at the center, and the TeO$_6$ octahedron is shaded maroon. (b) Crystal structure viewed along the $c$ axis. Magnetic Cu$^{2+}$ ions with spin-1/2 form a uniform square lattice in the $c$ plane. Dotted blue lines denote the chemical unit cell. (c) Exchange interactions $J_1$ and $J_2$ in the $ab$ plane.}
\label{fig1}
\end{figure}

In this short note, we report the results of the magnetic and specific heat measurements on a double perovskite, Sr$_2$CuTeO$_6$. This compound crystallizes in the tetragonal structure with the space group $I4/m$, as shown in Fig.\,\ref{fig1}\,\cite{Iwanaga,Reinen}. The crystal structure consists of CuO$_6$ and TeO$_6$ octahedra, shaded blue and maroon, respectively, which are linked sharing their corners. All the CuO$_6$ octahedra are elongated along the crystallographic $c$ axis owing to the Jahn-Teller effect. Consequently, the hole orbitals $d_{x^2{-}y^2}$ of Cu$^{2+}$ ions with spin-1/2 are spread in the $ab$ plane, in which Cu$^{2+}$ ions form a uniform squre lattice, as shown in Fig.\,\ref{fig1}(b). This leads to the strong superexchange interaction in the $ab$ plane and the weak superexchange interaction between the $ab$ planes. Thus, it is expected that Sr$_2$CuTeO$_6$ can be magnetically described as a quasi-two-dimensional $S\,{=}\,1/2$ $J_1\,{-}\,J_2$ SLHAF [Fig.\,\ref{fig1}(c)].
\begin{figure}
\begin{center}
\scalebox{0.4}{
\includegraphics{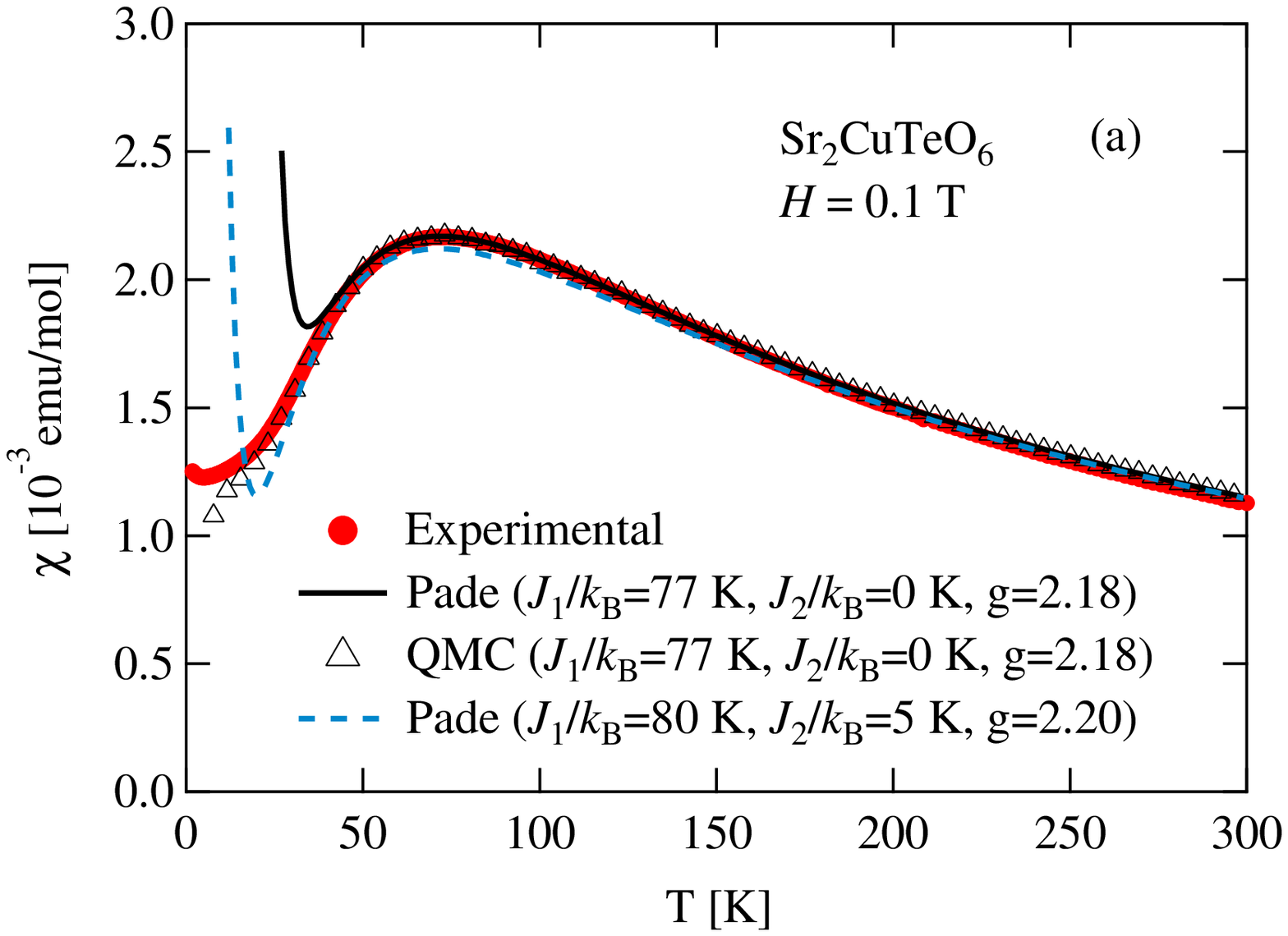}
}
\scalebox{0.4}{
\includegraphics{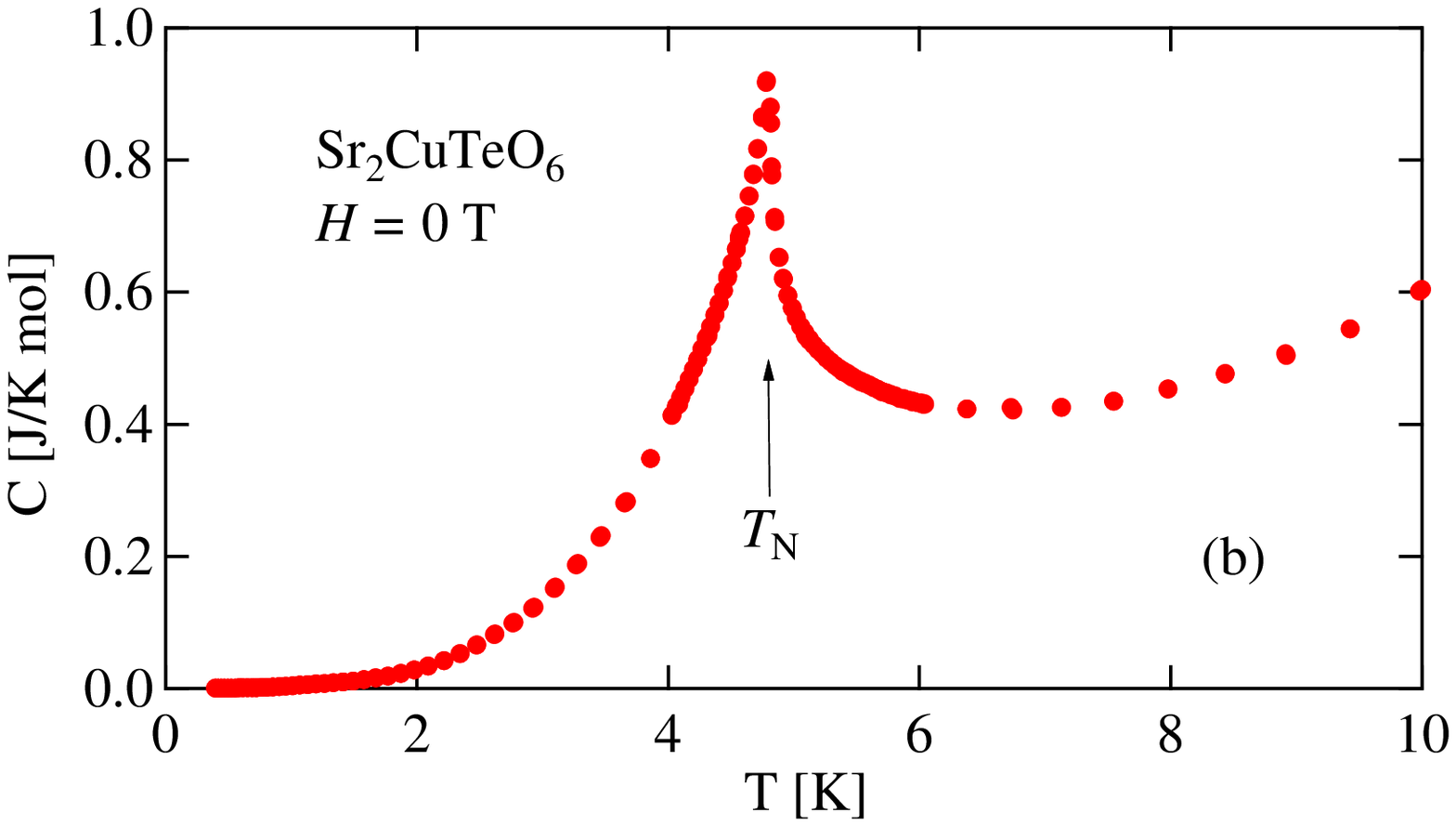}
}
\vspace{-3mm}
\end{center}
\caption{(Color online) (a) Temperature dependence of magnetic susceptibility $\chi$ in Sr$_2$CuTeO$_6$ measured at $H\,{=}\,0.1$ T (circular symbols). The solid and dashed lines show the susceptibilities calculated by the Pad\'{e} approximation with $(J_1/k_{\rm B}, J_2/k_{\rm B}, g)\,{=}\,(77\,{\rm K},0\,{\rm K}, 2.18)$ and $(80\,{\rm K},5\,{\rm K}, 2.20)$, respectively. The susceptibility calculated by the quantum Monte Carlo (QMC) method with $(J_1/k_{\rm B}, J_2/k_{\rm B}, g)\,{=}\,(77\,{\rm K},0\,{\rm K}, 2.18)$ is shown by triangular symbols. 
(b) Low-temperature specific heat in Sr$_2$CuTeO$_6$ measured at zero magnetic field.}
\vspace{-7mm}
\label{fig2}
\end{figure}

Figure\,\ref{fig2}(a) shows the magnetic susceptibility of Sr$_2$CuTeO$_6$ powder as a function of temperature. Our susceptibility data is consistent with that reported in Ref.\,\citen{Iwanaga}. With decreasing temperature, the susceptibility exhibits a rounded maximum at $T_{\rm max}\,{\simeq}\,73$ K and decreases. This behavior is characteristic of two-dimensional SLHAFs\,\cite{Rosner,de-Jongh,Kim}, and is common to the susceptibilities in A$_2$CuMO$_6$ with A\,=\,Ba, Sr and M\,=\,W, Mo, Te\,\cite{Iwanaga,Todate1,Vasala}.

Figure\,\ref{fig2}(b) shows the low-temperature specific heat in Sr$_2$CuTeO$_6$ measured at zero magnetic field. A sharp ${\lambda}$-like anomaly indicative of the magnetic ordering is observed at $T_{\rm N}\,{=}\,4.8$ K. This N\'{e}el temperature is much lower than those observed in isostructural A$_2$CuMO$_6$  with A\,=\,Ba, Sr and M\,=\,W, Mo, in which $T_{\rm N}\,{=}\,24-28$ K\,\cite{Todate2,Vasala}. 
The strong suppression of magnetic ordering in Sr$_2$CuTeO$_6$ should be ascribed to the weakness of the interlayer exchange interaction or to the strong spin frustration in the square lattice.

\begin{figure}[b]
\begin{center}
\scalebox{0.45}{
\includegraphics{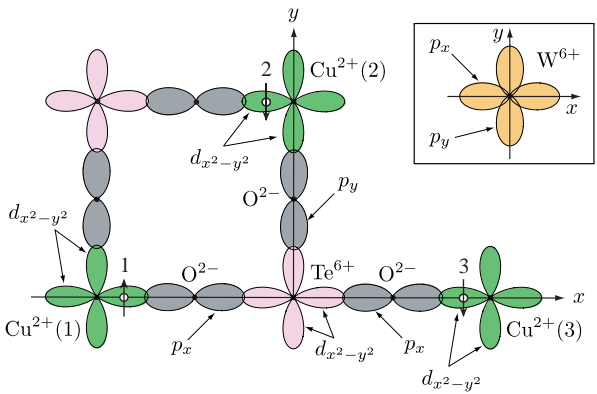}
}
\vspace{-2mm}
\end{center}
\caption{(Color online) Illustrations of orbital configurations related to superexchange interactions in A$_2$CuMO$_6$ through Cu$^{2+}$\,$-$\,O$^{2-}$\,$-$\,M$^{6+}$\,$-$\,O$^{2-}$\,$-$\,Cu$^{2+}$ for M\,=\,Te. The inset shows the configurations of the filled outermost $p$ orbitals in M\,=\,W and Mo.
}
\vspace{0mm}
\label{fig3}
\end{figure}

Before estimating the exchange constants from the susceptibility data, we discuss the superexchange interactions $J_1$ and $J_2$ in terms of the Kanamori theory.\cite{Kanamori} In the isostructural Ba$_2$CuWO$_6$, the antiferromagnetic $J_2$ interaction is dominant and the next-nearest-neighbor spins form the antiferromagnetic state below $T_{\rm N}\,{=}\,28$\,K\,\cite{Todate2}. This finding suggests that the exchange path Cu$^{2+}$\,$-$\,O$^{2-}$\,$-$\,M$^{6+}$\,$-$\,O$^{2-}$\,$-$\,Cu$^{2+}$ is dominant in A$_2$CuMO$_6$. Figure\,\ref{fig3} illustrates the orbital configurations in Sr$_2$CuTeO$_6$. For simplification, we assume that Cu$^{2+}$\,$-$\,O$^{2-}$\,$-$\,M$^{6+}$ is a straight line, although it is actually a zigzag line, as shown in Fig.\,\ref{fig1}(b).
The sign of the superexchange interaction strongly depends on the filled outermost orbital, which is the $4d$ orbital for Sr$_2$CuTeO$_6$. We consider the superexchange interaction between hole spins on the $d_{x^2{-}y^2}$ orbitals of Cu$^{2+}$ ions. The superexchange interaction $J_1$ between Cu$^{2+}$(1) and Cu$^{2+}$(2) is based on the following perturbation process. 
(i)~Hole 1 with up spin on Cu$^{2+}$(1) is first transferred to the $p_x$ orbital of O$^{2-}$, 
which is combined with the $d_{x^2{-}y^2}$ orbital of Te$^{6+}$ to form a molecular orbital. 
(ii)~Hole 2 on Cu$^{2+}$(2) is also transferred to other molecular orbital composed of the $p_y$ orbital of O$^{2-}$ and the $d_{x^2{-}y^2}$ orbital of Te$^{6+}$. 
In this case, the two hole spins on the same $d_{x^2{-}y^2}$ orbital of Te$^{6+}$ must be antiparallel owing to the Pauli principle. 
(iii)~The two holes are transferred back to the $d_{x^2{-}y^2}$ orbitals of two Cu$^{2+}$ ions. 
Consequently, an antiferromagnetic superexchange interaction takes place between Cu$^{2+}$(1) and Cu$^{2+}$(2). A similar perturbation process is also applicable to the superexchange interaction between Cu$^{2+}$(1) and Cu$^{2+}$(3). Thus, $J_2$ becomes antiferromagnetic. There are two Cu$^{2+}$\,$-$\,O$^{2-}$\,$-$\,Te$^{6+}$\,$-$\,O$^{2-}$\,$-$\,Cu$^{2+}$ paths for $J_1$, whereas for $J_2$, the exchange path is single. The contributions of the paths to the superexchange interaction should be almost the same. Thus, if the other exchange paths are negligible, the condition $J_2/J_1\,{\simeq}\,1/2$, which is in the critical region of the quantum disordered ground state, should be realized in Sr$_2$CuTeO$_6$. 
On the other hand, for M\,=\,W and Mo, the $p_x$ and $p_y$ orbitals of M$^{6+}$ are orthogonal to each other (inset of Fig.\,\ref{fig3}). Hence, two hole spins 1 and 2 on the $p_x$ and $p_y$ orbitals, respectively, must be parallel owing to the Hund rule. This leads to the ferromagnetic superexchange interaction $J_1$ between Cu$^{2+}$(1) and Cu$^{2+}$(2), while $J_2$ becomes antiferromagnetic as in the case of M=Te.

Here, we estimate the exchange parameters $J_1$ and $J_2$ of Sr$_2$CuTeO$_6$ from the susceptibility data by the [5, 5] Pad\'{e} approximation using the result of high-temperature expansion up to tenth-order of ${\beta}\,{=}\,1/k_{\rm B}T$.\cite{Rosner} We also calculate the susceptibility for $J_2\,{=}\,0$ case by the quantum Monte Carlo (QMC) method for the $12\,{\times}\,12$ site square cluster.\cite{Albuquerque} From the above discussion on the superexchange interactions, we can deduce that $J_1$ and $J_2$ interactions are both antiferromagnetic and $J_1\,{>}J_2$. We calculated susceptibilities varying $J_2/J_1$, and compared the results to the experimental susceptibility. It was found that $J_1$ is much larger than $J_2$ in Sr$_2$CuTeO$_6$ in contrast to the case of M\,=\,W and Mo. We show the susceptibilities calculated by the Pad\'{e} approximation with $(J_1/k_{\rm B}, J_2/k_{\rm B}, g)\,{=}\,(77\,{\rm K},0\,{\rm K}, 2.18)$ and $(80\,{\rm K},5\,{\rm K}, 2.20)$, respectively, and that calculated by the QMC method with $(J_1/k_{\rm B}, J_2/k_{\rm B}, g)\,{=}\,(77\,{\rm K},0\,{\rm K}, 2.18)$, which coincides with the result obtained by the Pad\'{e} approximation for $T\,{>}\,40$\,K. As shown by the solid line and triangular symbols in Fig.\,\ref{fig2}(a), the susceptibility is approximately represented in terms of a simple square-lattice antiferomagnetic Heisenberg model ($J/k_{\rm B}\,{=}\,77$\,K) without the next-nearest-neighbor interaction $J_2$. From the analysis of the susceptibility, exchange parameters in Sr$_2$CuTeO$_6$ are estimated as $J_1/k_{\rm B}\,{\simeq}\,80$\, K and $J_2/J_1\,{<}\,0.07$.
This result indicates that a superexchange path other than Cu$^{2+}$\,$-$\,O$^{2-}$\,$-$\,Te$^{6+}$\,$-$\,O$^{2-}$\,$-$\,Cu$^{2+}$ path, e.g., Cu$^{2+}$\,$-$\,O$^{2-}$\,$-$\,O$^{2-}$\,$-$\,Cu$^{2+}$ path, also makes an important contribution to $J_1$ interaction. For the accurate evaluation of the exchange parameters in Sr$_2$CuTeO$_6$, further experiments, such as a dispersion measurement, are necessary.


\begin{acknowledgment}
This work was supported by a Grant-in-Aid for Scientific Research (A) from the Japan Society for the Promotion of Science. 
Our QMC calculations were carried out using the ALPS application.\cite{Albuquerque}
\end{acknowledgment}

\end{document}